# Data Tastes Better Seasoned:

*Introducing the ASH Family of Hashing Algorithms*

D.J. Capelis
January 11th, 2005
California, United States
Senior, Technology High School
seasonedpaper@djc.people.inodetech.com

**Abstract**

Over the recent months it has become clear that the current generation of cryptographic hashing algorithms are insufficient to meet future needs. The ASH family of algorithms provides modifications to the existing SHA-2 family. These modifications are designed with two main goals: 1) Providing increased collision resistance. 2) Increasing mitigation of security risks post-collision. The unique public/private sections and salt/pepper design elements provide increased flexibility for a broad range of applications. The ASH family is a new generation of cryptographic hashing algorithms.

## Background

Over the recent month, the security and cryptography communities have been discussing the collisions found in MD5, MD4, RIPEMD and HAVEL-128 by a team lead by Wang[1], and the collision found in SHA-0 (Secure Hashing Algorithm) by a team led by Joux[2] as well as his work with other iterated algorithms. Their findings spurred the recent publication of the paper "MD5 To Be Considered Harmful Someday"[3] by Dan Kaminsky which proposed a possible attack vector on the MD5 hashing algorithm using a collision discovered by Wang et al.

These discoveries leave the United States' National Security Agency's (NSA) SHA-1/224/256/368/512 as the only mainstream algorithms for which cryptographers have yet to find collisions. These are also the only algorithms specified as the SHS (Secure Hash Standard) under FIPS (Federal Information Processing Standard) 180-2.[4]

However, even these algorithms are showing weaknesses. In late August 2004 the National Institute for Standards and Technology (NIST, the same body which releases the SHS and the FIPS) applauded the recent research and announced the phase out of SHA-1 by 2010[5]. Though these algorithms have remained strong and cryptanalysis on the newer algorithms have turned up no significant problems, it has become apparent that the current class of algorithm is rapidly becoming insufficient for future security needs.

The ASH algorithms attempt to fill this need. Offering increased flexibility and security, the ASH algorithms are intended to provide modifications that will enhance overall security. Due to the security record of the SHA family and the availability of the new algorithms released in 2002, SHA-256 and SHA-512 have been chosen as the base algorithms for ASH-1 and ASH-2.

# Contents
- Objectives
- Hash Format
- The Appendable Cascade Problem
- ASH Data Restructuring
- Proper Seasoning of Data
- Secret Seasoning
- Implementing a Hashing Utility or API Based on ASH
- ASH in Protocols

# Objectives

The ASH algorithms avoid the necessity to design new algorithms from the ground up. Instead, the ASH algorithms are based on an existing algorithm. For ASH-1 and ASH-2, algorithms in the SHA family have been selected. ASH is designed to treat the actual hash function as a black-box that can easily be changed as newer and improved hash functions are created.

ASH reorders and modifies data as it is fed into the hash function, this results in increased security and offers improved mitigation of collisions when they do take place. The goal is to increase resistance, not provide an absolute solution to any particular problems as to do so would be impossible or infeasible. ASH's series of modifications are not SHA specific and can be applied to any iterative hashing function.

For the first several algorithms, ASH is merely SHA with alterations to the datastream as it enters the hashing function. (Hence the similarity in the names of the algorithms.) However, the modifications increase the flexibility of the algorithms immeasurably. A pepper (explained later) is also used. The pepper allows for multiple dynamic sections to be generated. Each section can be verified to increase the security of the system. (This is discussed the the "Secret Seasoning" section.)

# Hash Format

**ASH-1 Hash Format:**

*Total Size: 1024 Bits (256 + 256 + 512)*

| SHA-256 Hash | SHA-256 Hash | Pepper Data (512 Bits) |
|---|---|---|
| Static Section | Dynamic Section ||

**ASH -2 Hash Format:**

*Total Size: 2048 Bits  (512 + 512 + 1024)*

| SHA-512 Hash | SHA-512 Hash | Pepper Data (1024 Bits) |
|---|---|---|
| Static Section | Dynamic Section ||

# The Appendable Cascade Problem

The attack vector detailed by Kaminsky in "MD5 To Be Considered Harmful Someday" relies on the Appendable Cascade Problem. This problem is based on the fact that MD5, like all iterative hash functions, is basically a block algorithm. Blocks are fed into the algorithm in distinct pieces. For example, imagine a file with 5 blocks:

| A | B | C | D | E |
|---|---|---|---|---|

Now, imagine an attacker generates A' so that Hash(A') = Hash(A). After A and A' collide, the value of the hash is identical. Further blocks can be fed into the algorithm and as long as the blocks continue to have identical hash values, the resulting hash for both files will be the same, even though the file as a whole is different. Mathematically:

$$Hash( A + B+ C+ ...) = Hash( A' + B + C + ...)$$

Without this problem, the attack scenario described in the work by Kaminsky (MD5 To Be Considered Harmful Someday) would be much less feasible.

## ASH Data Restructuring

To reduce the feasibility of exploiting this vulnerability, ASH is designed to be "Appendable Cascade" resistant. ASH accomplishes this goal by simply restructuring the data. For example, the base algorithm for ASH-1 (SHA-256) uses a blocksize of 512 bits. Conceptually, ASH splits these blocks in half, like so:

| A | | B | | C | | D | | E | |
|---|---|---|---|---|---|---|---|---|---|
| A1 | A2 | B1 | B2 | C1 | C2 | D1 | D2 | E1 | E2 |

Now the blocks can be renumbered as separate 256-bit pieces and reordered and combined back into blocks: (The numbers are bolded and italicized to more clearly show the reordering process.

| 1 | 2 | 3 | 4 | 5 | 6 | 7 | 8 | 9 | 10 |
|---|---|---|---|---|---|---|---|---|---|
| **1** | *6* | **2** | *7* | **3** | *8* | **4** | *9* | **5** | *10* |
| 1 & 6 | | 2 & 7 | | 3 & 8 | | 4 & 9 | | 5 & 10 | |

So a collision (in which Hash(1' + 6') = Hash(1 + 6)) looks like:

| 1' & 6' | 2 & 7 | 3 & 8 | 4 & 9 | 5 & 10 |
|---|---|---|---|---|

If blocks are appended to the end of the datastream, hashes no longer collide after the reordering process:

| 1 & 7 | 2 & 8 | 3 & 9 | 4 & 10 | 5 & 11 | 6 & 12 |
|---|---|---|---|---|---|
| 1' & 7 | 2 & 8 | 3 & 9 | 4 & 10 | 5 & 11 | 6' & 12 |

Appendable Cascading is still quite possible, but the file must be restructured. This small caveat is what makes the algorithm stronger and significantly raises the bar for exploitation with ordered file formats. Also, a salt can be added to two files to ensure integrity.

If foul play is suspect, instead of openly transferring and comparing files, two computers can create a block of data (each computer produces half) and append this to the file. Then, each computer can create a new hash that should never match up if the files are different. (1 in $2^{128}$ chance for ASH-1, 1 in $2^{256}$ chance for ASH-2)

## Proper Seasoning of Data

ASH doesn't automatically apply a salt, instead, a pepper is applied: After the re-ordering process of ASH, the blocks are fed into the hashing algorithm. However, in ASH, hashing can be done with more than one processing unit. This is because simultaneously these same blocks are being XOR'd with a pepper and a second hash is being created. A pepper is simply one block of random data. It's name steams from the purpose of the pepper which is to pepper around bit-changes, so to speak. To verify a hash, a computer simply uses the pepper data embedded in the hash format.

In comparison with the peppering method, a salt adds one block onto the end of the datastream. A pepper provides increased security because a pepper causes bit-changes through the entire datastream, these changes make it harder to generate collisions. A pepper also allows for multiple and secret hashes which each can be verified independently.

## Secret Seasoning

As an example, pretend a college student from Finland creates a really cool operating system. (This seems to happen every once in awhile.) For easy reference, call this student Linus. In this hypothetical situation, Linus wishes to release the next version of his operating system and wants to create a hash so end-users can ensure that the no one has inserted malicious code.

Let's assume Linus uses ASH to publish a hash. He would also want to generate a second hash and store it in a secure place. This way, he can keep doing what he loves, developing his operating system, without having to worry about keeping backups. All Linus has to do is upload the release to FTP and let others mirror it.

Imagine that 20 years in the future, the archive is compromised and Linus is looking to replace the file. He has long since deleted the old release. The attacker has replaced the release with a version that has been modified to transmute the computer into an ordinary toaster oven[6]. Unfortunately, this version has been constructed to verify correctly with the original published hash.

Within hours, Linus has several offers from people who happen to have a backup of the release. However, some offers of help seem decidedly suspicious. Linus would want to verify that the releases that people are sending to him is the same release he made twenty years ago.

All Linus needs to do is verify these backups with his secure hash. If the hash shows that the backups will match the original hash, Linus knows it's a good copy. This is because it is impossible for an attacker to pre-create data that would verify correctly if he does not know the pepper. This is the concept of secret seasoning. A secret hash can be kept to later verify a file even if a malicious hash is created or a full break in base algorithm arises that makes collisions trivial.

## Implementing a Hashing Utility or API Based on ASH

ASH permits files to have multiple hashes. (ASH-1 allows $2^{256}$, ASH-2 allows $2^{512}$) This means that each time a hash is created, it will likely be different, even if the files are the same. And further, verification mode is separate from the creation mode. To properly verify a hash, both hashes must use the same pepper. It follows that the process of creating a hash is separate from the process of verifying a hash. Hash creation uses a random pepper, hash verification must read the pepper from the hash against which a datastream is being verified.

## ASH In Protocols

ASH's pepper creates a unique circumstance for use in protocols. To create a pepper, computers should all generate a block of data and XOR them together. If any computer generates a block of truly random data, then the pepper will be completely random. These same principles of operation allow a one-time pad to produce unbreakable encryption.

Multiple peppers allow for greatly increased flexibility for hashes in encryption protocols. For example, ASH by nature provides innate challenge-response authentication because an attacker will be unable to generate a hash without obtaining the data. This is one of many cases where creative use of ASH's peppers results in increased flexibility.

## Conclusions

ASH can be implemented immediately. This provides increased security and flexibility. Further work and research in this area is encouraged. Comments on ASH can be sent via e-mail.

## Licensing

This paper establishes prior art for any patents which attempt to claim ownership over ASH technology. All methods, algorithms and technology described in this paper may be used without acknowledgment or compensation. For the actual paper, the current copyright statement is as follows: "Copyright D.J. Capelis 2005. All rights reserved."

This paper will be re-released under a creative commons license after the organization releases their licenses for science related publications. This revision is a work in progress and will only be available from the Eprint cryptography archive. Fair use rights should be interpreted liberally and this paper will be relicensed to allow for increased distribution as soon as acceptable licenses are available.

Further information on licensing issues can be found here: http://science.creativecommons.org. This paper will likely be re-released under one of the licenses created by the project.